\documentclass[preprint]{aastex631}

\submitjournal{Universe}

\shorttitle{2021 Periastron of PSR B1259-63}
\shortauthors{Chang et al.}

\graphicspath{{./}}

\begin{document}

\title{\textit{Fermi}-LAT Observation of PSR B1259-63 during Its 2021 Periastron Passage}

\correspondingauthor{Zhi Chang}
\email{changzhi@ihep.ac.cn}

\author[0000-0003-4856-2275]{Zhi Chang}
\affiliation{Key Laboratory of Particle Astrophysics,
    Institute of High Energy Physics,
    Chinese Academy of Sciences,
    Beijing 100049, China
}

\author{Shu Zhang}
\affiliation{Key Laboratory of Particle Astrophysics,
    Institute of High Energy Physics,
    Chinese Academy of Sciences,
    Beijing 100049, China
}

\author{Yu-Peng Chen}
\affiliation{Key Laboratory of Particle Astrophysics,
    Institute of High Energy Physics,
    Chinese Academy of Sciences,
    Beijing 100049, China
}

\author{Long Ji}
\affiliation{School of Physics and Astronomy,
	Sun Yat-Sen University, 
	Zhuhai 519082, China
}

\author{Ling-Da Kong}
\affiliation{Key Laboratory of Particle Astrophysics,
    Institute of High Energy Physics,
    Chinese Academy of Sciences,
    Beijing 100049, China
}
\affiliation{University of Chinese Academy of Sciences, 
	Beijing 100049, China
}

\author{Peng-Ju Wang}
\affiliation{Key Laboratory of Particle Astrophysics,
    Institute of High Energy Physics,
    Chinese Academy of Sciences,
    Beijing 100049, China
}
\affiliation{University of Chinese Academy of Sciences, 
	Beijing 100049, China
}

\begin{abstract}
PSR B1259-63 is a $\gamma$-ray binary system,
where the compact object is a pulsar.
The system has an orbital period of 1236.7 days and shows peculiar $\gamma$-ray flares
(in 100\,MeV--300\,GeV) after its periastron time.
We analyzed the \textit{Fermi}-LAT observation of PSR B1259-63
during its latest periastron passage,
as well as its previous three periastrons.
The bright GeV flares started about 60 days after the periastron epoch in 2021.
This delay is larger than that around the 2017 periastron and much larger than earlier periastrons.
The delay of the GeV flux peak time in each periastron passage is apparent in our results.
We discussed the possible origin of this delay
and made a prediction of the GeV flux peak time in next periastron passage,
based on observation of the previous delays.
\end{abstract}


\section{Introduction} \label{sec_intro}

Gamma-ray binaries are systems
composed of a massive star in orbit with a compact object
and are characterized by broad non-thermal emission peaking (in $\nu F_\nu$)
at energies above 1\,MeV \citep{2013A&ARv..21...64D, 2017A&A...608A..59D}.
Among the $\gamma$-ray binaries,
PSR B1259-63 \citep{1997ApJ...477..439T} and PSR J2032 + 4127 \citep{2009Sci...325..840A}
are the only two for which the nature of compact object is known
(both are pulsars).
In PSR B1259-63/LS 2883,
the compact object is a non-recycled, spin-down powered radio pulsar,
with a period of 47.76\,ms, orbiting a Be star (LS 2883) with a period of $\sim$1236.7\,days
in a highly eccentric orbit ($e\sim0.87$)
\citep{1992ApJ...387L..37J, 2014MNRAS.437.3255S}.
The pulsar is at a distance of $\sim$0.9\,AU
from its companion star at periastron \citep{1992ApJ...387L..37J}.
The pulsar will cross the equatorial disk of the Be type companion twice in one orbit,
as the orbital plane of the pulsar is thought to be inclined ($\approx$10$^\circ$)
with respect to this equatorial disk \citep{1995MNRAS.275..381M}.

The multi-wavelength emission of PSR B1259-63 is usually thought to result from
the bow-shaped shock interaction between the relativistic pulsar wind and the stellar wind.
The radio, X-ray and TeV radiation from PSR B1259-63
show a similar behaviour, exhibiting a two-peak light curve
\citep{1992ApJ...387L..37J, 2015MNRAS.454.1358C, 2020MNRAS.497..648C, 2020A&A...633A.102H}.
Significant GeV $\gamma$-ray emission above 100\,MeV was detected
by \textit{Fermi}-LAT around the last four periastron passages,
and it shows a very different behavior to the other energy bands.
The GeV emission started approximately tens of days after the periastron,
and lasted more than one month.
Multiple significant flares are seen in the detected GeV emission,
and show very different properties on different time scales
\citep{2018RAA....18..152C, 2018ApJ...863...27J, 2018ApJ...862..165T}.
\citet{2018RAA....18..152C} also reported a time delay of the main GeV flare
by comparing the three previous periastron passages
(2010, 2014 and 2017).
The most recent multi-wavelength observations
which covered a period of more than 100 days around the 2021 periastron
show substantial differences from the previously observed passages \citep{2021Univ....7..242C}.
Besides the features in other energy bands,
such as the weaker X-ray flux during peaks and the never before observed third X-ray peak,
\citet{2021Univ....7..242C} also confirmed the delay of the GeV flare
reported by \citet{2018RAA....18..152C} in 2018.

Since the physical origin of both the flares and the time delay of the flares
observed in GeV energy band remains unknown,
here using the latest instrument response functions and more than 10 years of data,
we report on detailed \textit{Fermi}-LAT observations
of the latest periastron passage of PSR B1259-63 in February 2021,
and compare this to the previous periastron passages in 2010, 2014 and 2017.
The paper is structured as follows.
The observations and data analysis are described in Section \ref{sec_obs},
the results and discussions are shown in Section \ref{sec_results},
and we give our conclusions in Section \ref{sec_conclusions}.

\section{Observations and Data Analysis} \label{sec_obs}

The Large Area Telescope (LAT) on-board the \textit{Fermi} satellite
is an electron-positron pair production telescope
operating at energies from $\sim$100\,MeV to greater than $300$\,GeV \citep{2009ApJ...697.1071A}.
It has a large field of view (about 20\% of the sky) and
has been scanning the sky continuously since August 2008.
\textit{Fermi}-LAT observed PSR B1259-63 during its periastron period in 2010
\citep{2011ApJ...736L..11A, 2011ApJ...736L..10T},
2014 \citep{2015ApJ...811...68C},
2017 \citep{2018RAA....18..152C, 2018ApJ...863...27J, 2018ApJ...862..165T}
and 2021 \citep{2021ATel14399....1J, 2021ATel14540....1J, 2021ATel14612....1J, 2021Univ....7..242C}.

The analysis of \textit{Fermi}-LAT data was performed using the {\tt Fermitools}
package (version 2.0.8, released on 2021 January 21).
\footnote{\url{https://fermi.gsfc.nasa.gov/ssc/data/analysis/software/}
(accessed on 1 December 2021)}
The Pass 8 {\tt SOURCE} event class
({\tt evtype  =  128})\footnote{\url{https://fermi.gsfc.nasa.gov/ssc/data/analysis/documentation/Pass8\_usage.html} (accessed on 1 December 2021)}
was included in the analysis using the {\tt P8R3\_SOURCE\_V3} instrument response functions (IRFs).
All $\gamma$-ray photons used for this analysis were
within an energy range of 100\,MeV--300\,GeV
and within a circular region of interest (ROI) of 10$^\circ$ radius centered on PSR B1259-63.
Time intervals when the region around PSR B1259-63 was observed
at a zenith angle less than 90$^\circ$ were selected
to avoid the contamination from the earth limb $\gamma$-rays.
The Galactic and isotropic diffuse emission components as well as known $\gamma$-ray sources
within 15$^\circ$ of the ROI center based on the
\textit{Fermi}-LAT 10-year Source Catalog (4FGL-DR2) \citep{2020ApJS..247...33A, 2020arXiv200511208B}
were considered in our analysis.
The spectral  parameters were fixed to the catalog values,
except for the sources within 3$^\circ$ of PSR B1259-63,
for which the flux normalization was left free.
In the modeling of PSR B1259-63, a simple power-law ({\tt PowerLaw}) and
an exponentially cut-off broken power-law ({\tt PLSuperExpCutoff}, fixed index2 = 1)
were tested, respectively.
The definitions of both models can be found in {the} \textit{Fermi}-LAT support page.\footnote{\url{https://fermi.gsfc.nasa.gov/ssc/data/analysis/scitools/source_models.html}
(accessed on 1 December 2021)}
The significance of the sources was evaluated by the Test Statistic (TS).
This statistic is defined as TS  =  $-$2\,$\ln(L_{max,0}/L_{max,1})$,
where $L_{max,0}$ is the maximum likelihood value
for a model in which the source studied is removed (the ``null hypothesis''),
and $L_{max,1}$ is the corresponding maximum likelihood value with this source being incorporated.
The square root of the TS is approximately equal to the detection significance of a given source.

The precise time of four periastron passages of PSR B1259-63 used in our work are
MJD 55544.69378 (14 November 2010, 16:39:03 UTC),
MJD 56781.41830 (4 May 2014, 10:02:22 UTC),
MJD 58018.14283 (22 September 2017, 03:25:41 UTC) and
MJD 59254.86735 (9 February 2021, 20:48:59 UTC).
They are derived from the orbital ephemeris ($P_b$ = 1236.724526(6) days)
as reported in \citet{2014MNRAS.437.3255S}.

\section{Results and Discussion} \label{sec_results}

\begin{figure*}
    \centering\includegraphics[width=0.48\textwidth]{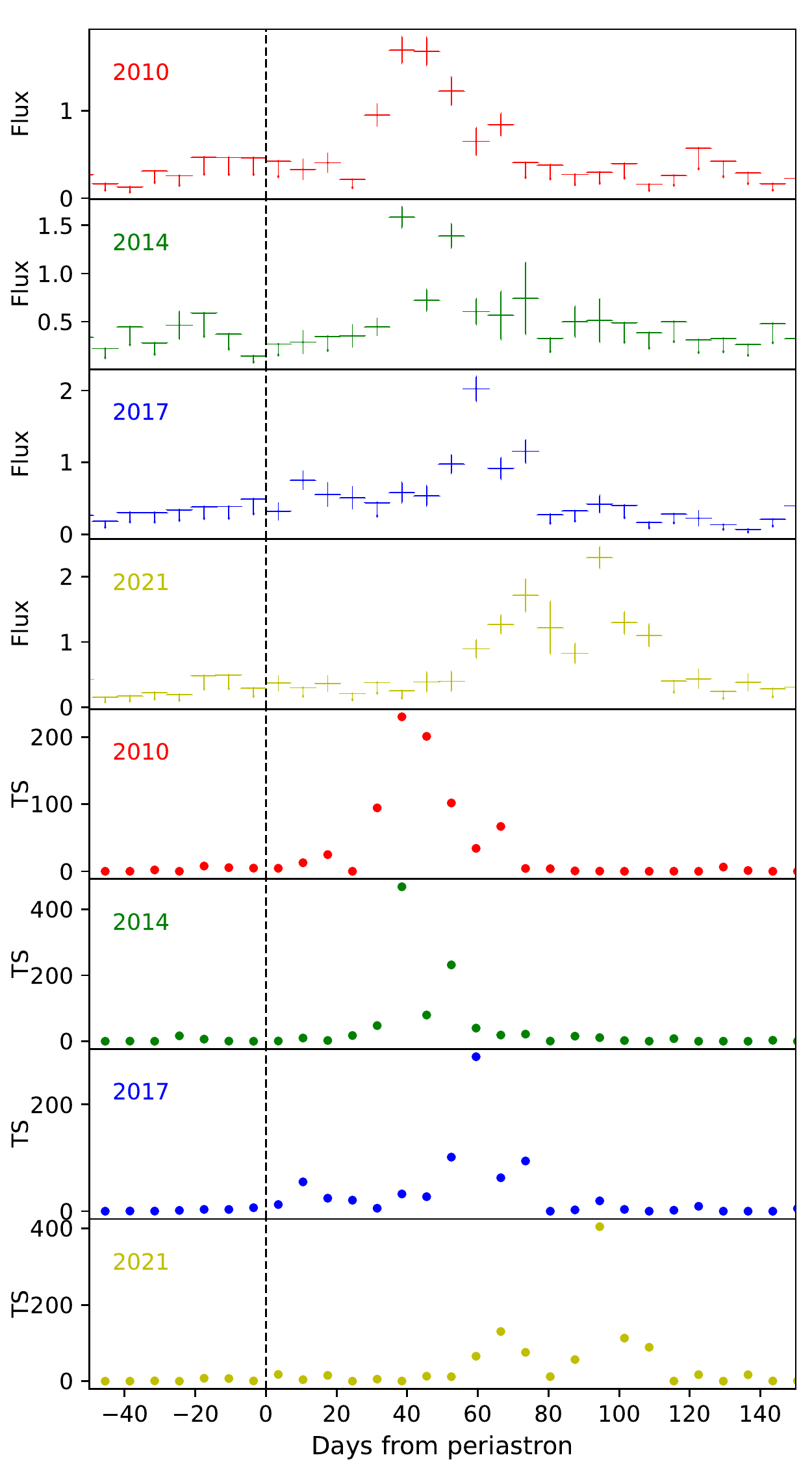}
    \centering\includegraphics[width=0.48\textwidth]{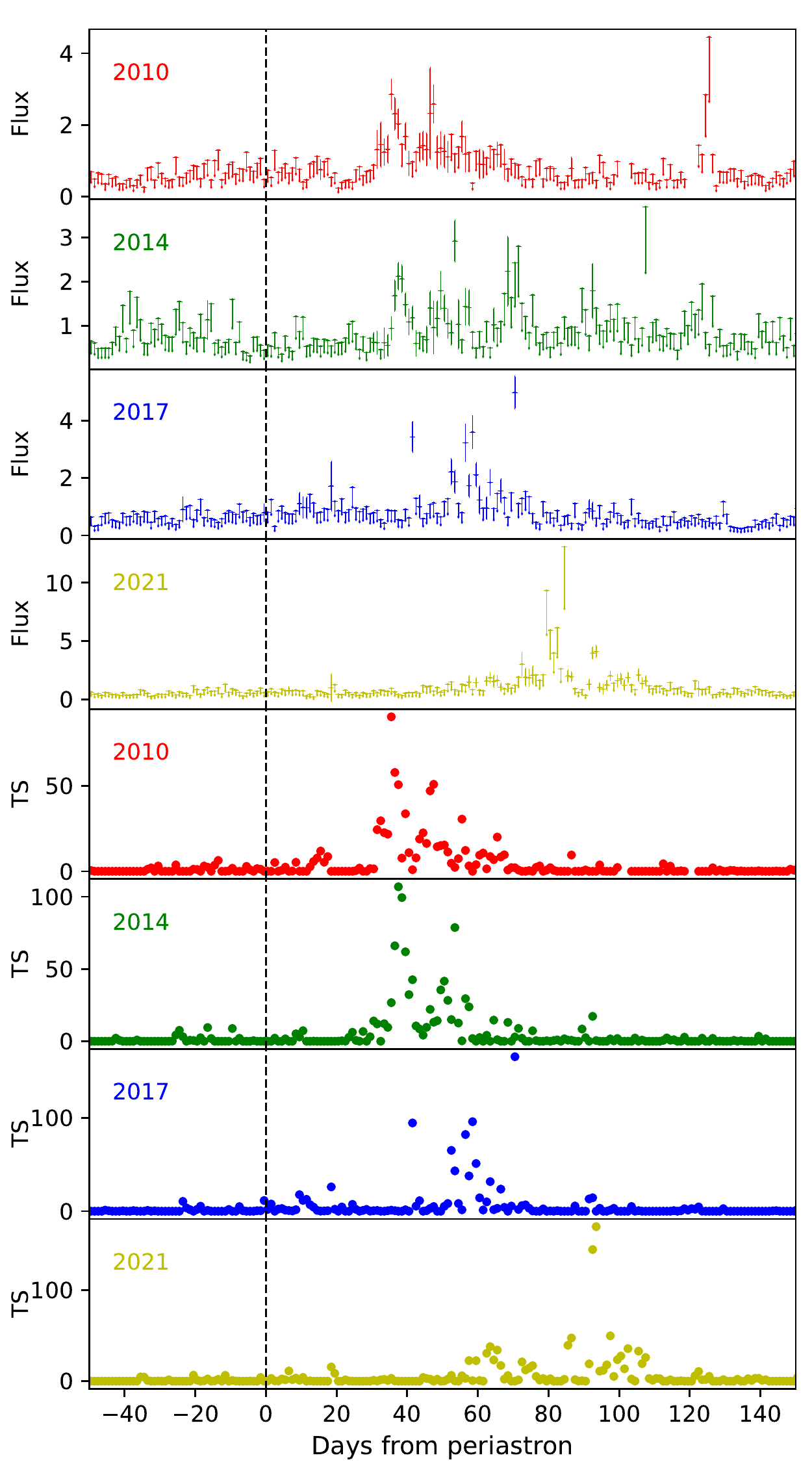}
    \caption{Weekly (left panels) and daily (right panels)
    light curves (top four panels, flux in unit 10$^{-6}$ ph cm$^{-2}$ s$^{-1}$)
    and TS values (bottom four panels)
    during \textit{Fermi}-LAT flares as observed in 2010 (red), 2014 (green) 2017 (blue)
    and 2021 (yellow), respectively.
    The dashed black vertical line indicates the periastron time.
    \label{fig_lc_daily_weekly}}
\end{figure*}

Using the binned likelihood analysis,
we produced the \textit{Fermi}-LAT weekly and daily light curves of PSR B1259-63
from 50 days before until 150 days after the 2021 periastron passage,
shown in Figure \ref{fig_lc_daily_weekly} (yellow points in both panels).
A 95\% confidence upper limit was calculated
when PSR B1259-63 was not significantly detected (TS $<$ 9).
The results are consistent with those reported in the 0.1--10\,GeV energy band
in \citet{2021Univ....7..242C}.
For a comparison to previous periastron passages,
we re-analyzed all the \textit{Fermi}-LAT data during its 2010, 2014 and 2017 periastron passages
with the latest IRFs and the latest software package.
All the \textit{Fermi}-LAT light curves of the four periastron passages
in 2010, 2014, 2017 and 2021 are showed in Figure \ref{fig_lc_daily_weekly}.
For each light curve,
significant GeV emission is mainly detected between 10 and 120 days after each periastron.

\begin{figure*}
	\centering\includegraphics[width = 0.95\textwidth]{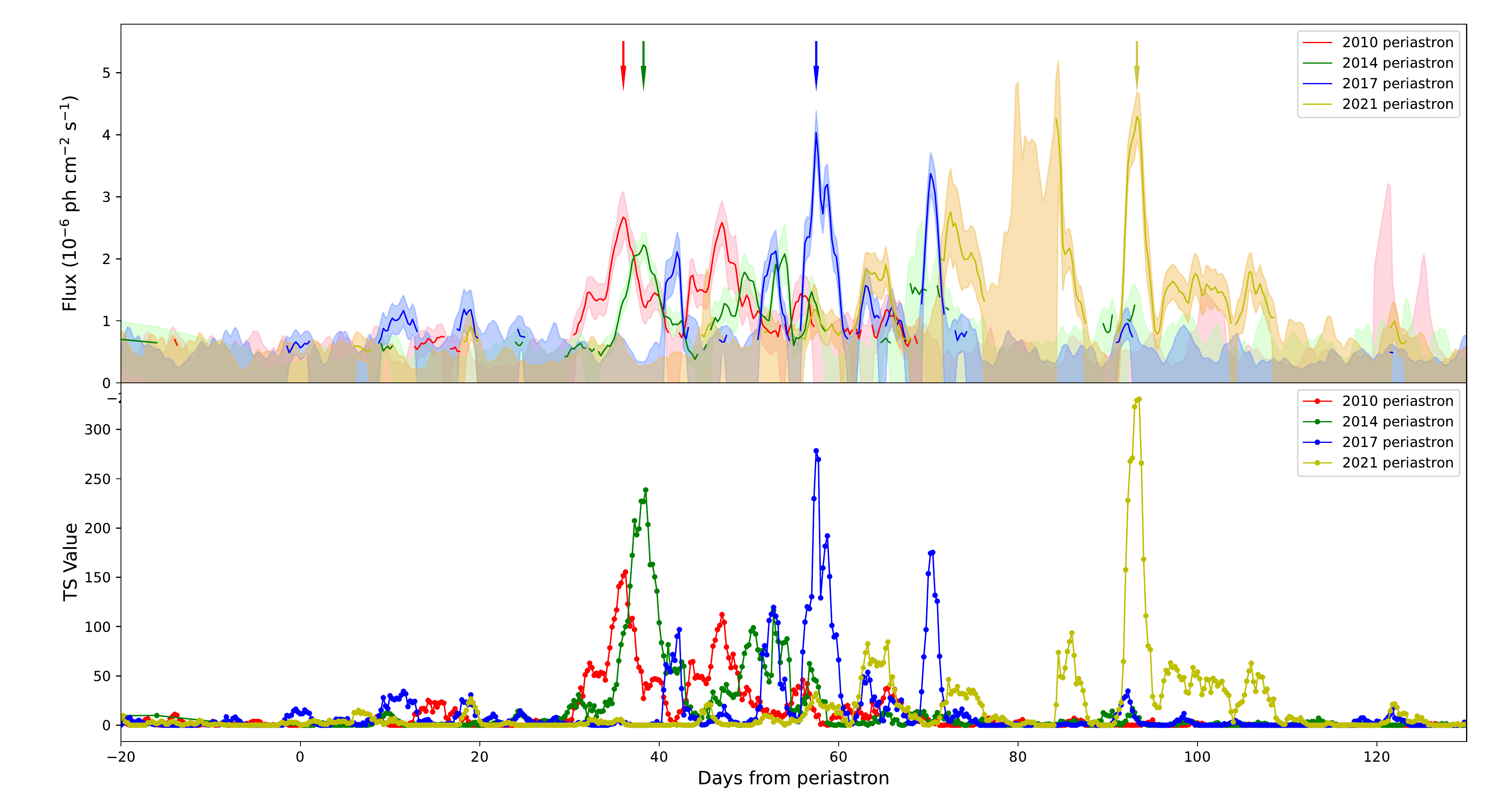}
	\caption{Smoothed light curves and TS values for flares observed by \textit{Fermi}-LAT in
    2010, 2014, 2017 and 2021 periastron passages of PSR B1259-63.
    The shaded areas show the statistical error zones or the upper limits
    (at 95\% confidence level, when whose TS value is less than 9) of the flux.
    The arrows in upper panel indicated the highest flux during each periastron.
    }
\label{fig_lc_sliding}
\end{figure*}

\begin{figure*}
\centering\includegraphics[width = 0.7\textwidth]{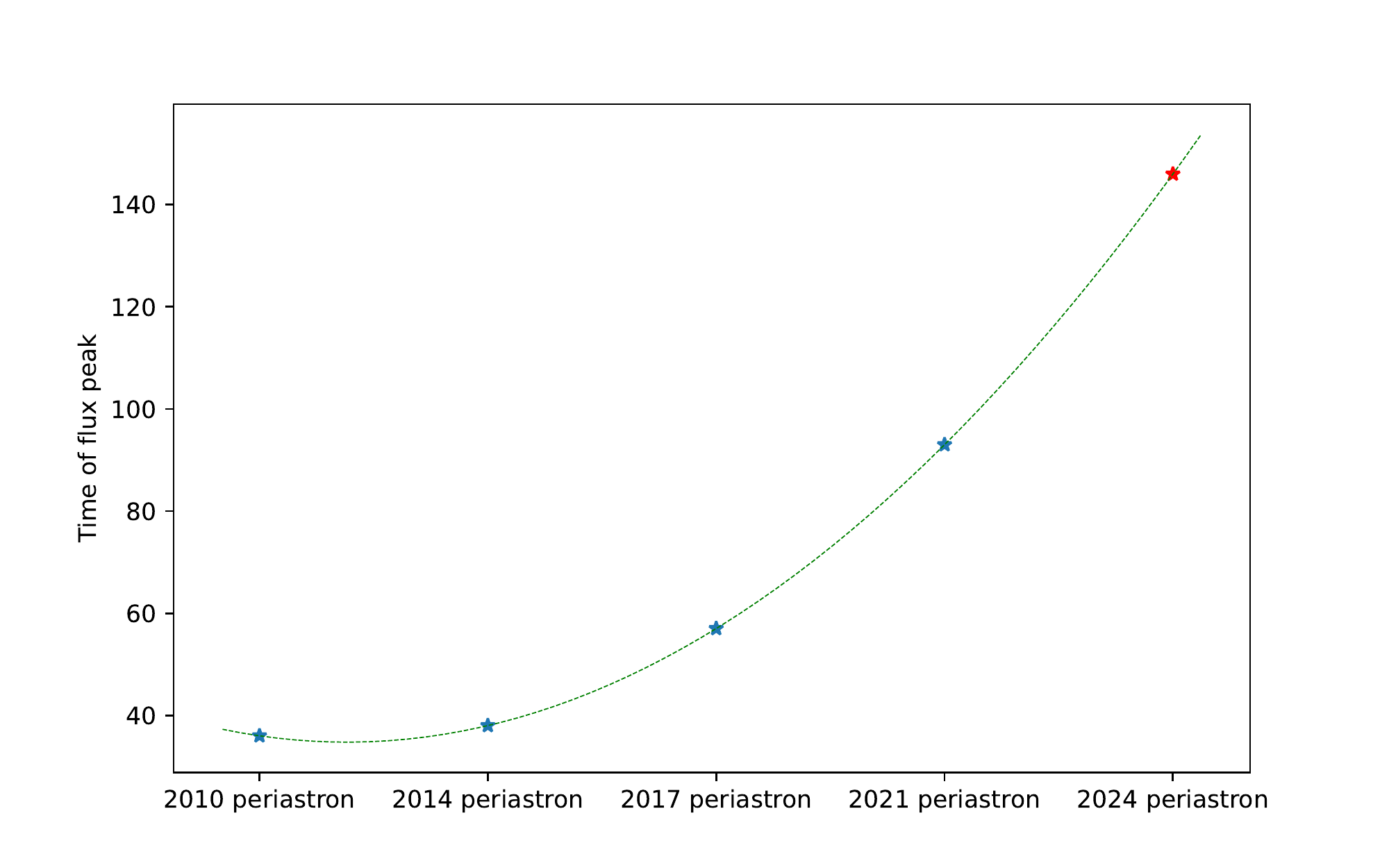}
	\caption{The delay in the time of the main GeV flue peak for each periastron.
	The peak is defined as the flux peak during each periastron passage shown in
	Figure \ref{fig_lc_sliding}. Sinusoidal function that is used to fit the data points.
	We predict the GeV flux peak time of next periastron passage in 2024,
	shown as a red star in the figure.}
\label{fig_delay}
\end{figure*}

In order to quantify the difference of the light curve profiles clearly
during these four periastron passages,
smoothed light curves were produced with the sliding windows technique
introduced in \citet{2015ApJ...811...68C}, 
which are shown in Figure \ref{fig_lc_sliding}.
We choose time windows of 2 days, whose starting times lag the previous ones by 6 h.
In this analysis, a binned likelihood analysis was performed in every window.
The spectral index of PSR B1259-63 was allowed to vary between 1.0 and 4.0.
The shaded areas indicate the statistical error of the flux.
Similarly to the weekly and daily light curves (shown in Figure \ref{fig_lc_daily_weekly}),
a 95\% confidence upper limit was calculated
when the source was not significant detected (TS $<$ 9).
Here we define the main peak of the GeV emission as
the highest flux in the smoothed light curve during each periastron passage,
shown in Figure \ref{fig_lc_sliding},
indicating with arrows with corresponding colors.
The time of the main GeV flux peak
is shown as the number of days after periastron in Figure \ref{fig_delay}.
It is apparent that the GeV flux peak occurs later each orbit
and the time delay is increasing between periastrons.

Judging from Figures \ref{fig_lc_daily_weekly} and \ref{fig_lc_sliding},
the light curves shows a complex behavior.
For all periastron passages the most significant GeV emission
(with very bright flux being detected) occurred more than 30 days after periastron,
although some flares were also significantly detected before 20 days after periastron.
The main GeV flux peak appeared later and later in the last four periastron passages,
as shown in Figure \ref{fig_delay},
occuring about 90 days after periastron in 2021.
This delay pattern is consistent with the trend reported in \citet{2018RAA....18..152C},
and it is not being observed in any other energy bands \citep{2021Univ....7..242C}.
Another feature of the light curves is that
there are multiple bright flares during each periastron passage.
These bright flares in each light curve do not show any pattern.

\begin{figure*}
\centering\includegraphics[width = 0.9\textwidth]{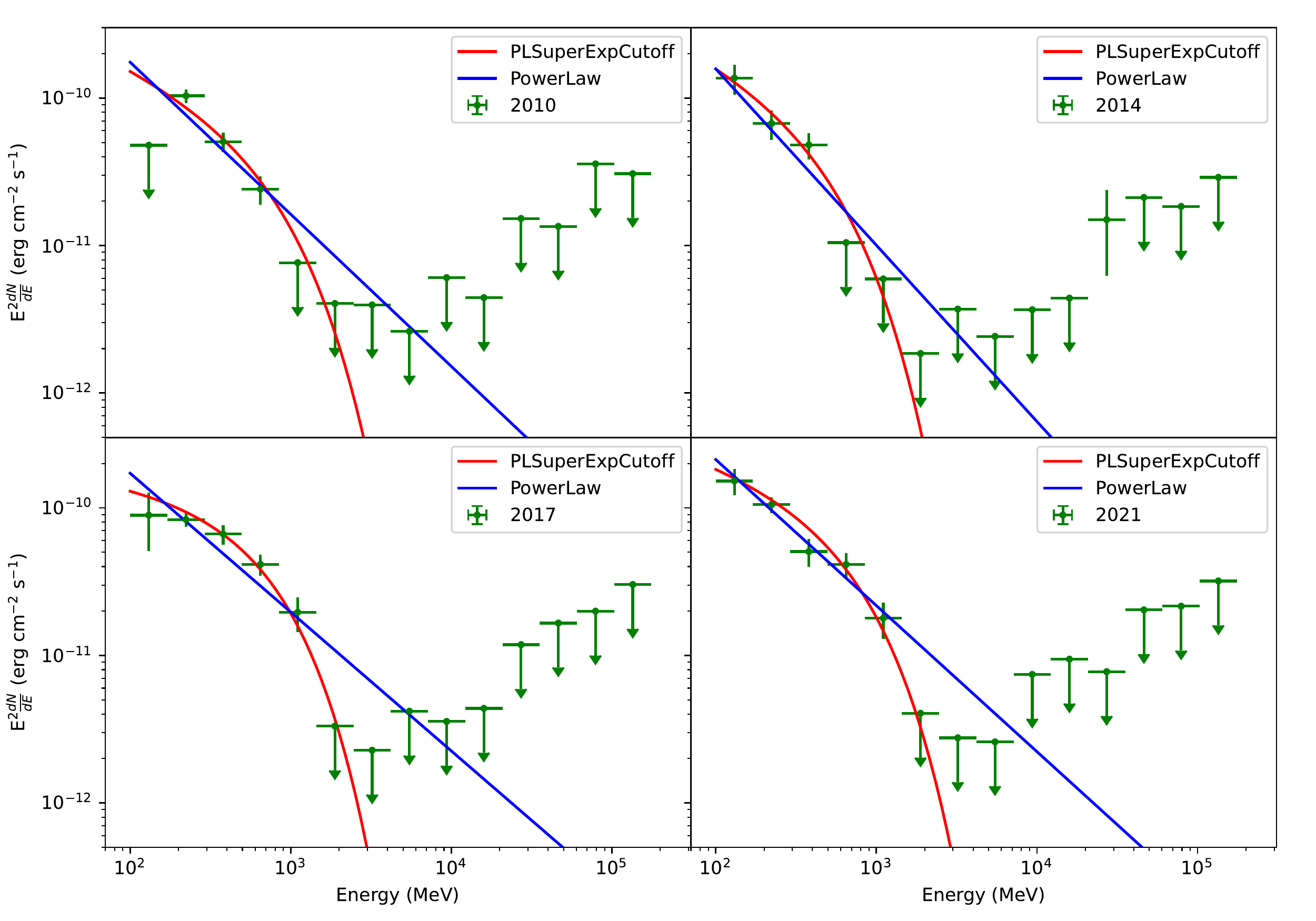}
	\caption{SEDs of PSR B1259-63 derived from binned likelihood analysis of \textit{Fermi}-LAT,
    between 0 and 120 days after each periastron.
    A simple power-law and a power-law with exponentially cut-off are used.
    The best-fit parameters of each model can be found in Table \ref{tab_model_pars}.
    }
\label{fig_spectrum}
\end{figure*}

The light curves in Figures \ref{fig_lc_daily_weekly} and  \ref{fig_lc_sliding}
suggest different behaviors in different periastron passages.
To study the spectral variability of the source,
we analyzed the spectrum of PSR B1259-63 between 0 and 120 days after each periastron.
Both a simple power-law and a power-law with exponentially cut-off model
were used in our analysis.
The resulting spectral energy distributions (SEDs) are shown in Figure \ref{fig_spectrum},
and the best-fit parameters of each model are summarized in Table \ref{tab_model_pars}.
The exponential cut-offs are significantly detected for the 2014, 2017 and 2021 periastron
(but not for the 2010 periastron) as $\Delta{\rm TS} > 9$
($\Delta$TS = $-2 \ln (L_{\rm PL}/L_{\rm CPL})$,
where $L_{\rm CPL}$ and $L_{\rm PL}$ are the maximum likelihood values for power-law models
with and without a cutoff. If $\Delta{\rm TS} > 9$, a cutoff is significantly preferred).
This is consistent with the results reported in
\citet{2020MNRAS.497..648C, 2021Univ....7..242C}.
Using a power-law with exponentially cut-off model,
the spectra index shows significant change between the 2014 and 2017 periastron.

\begin{table}[]
\caption{The best-fit parameters of the SED models.
	$\Delta$TS = $-2 \ln (L_{\rm PL}/L_{\rm CPL})$, detailed definition can be found in the text.}
\centering
\begin{tabular}{ccccccc}
\hline
       &\multicolumn{2}{c}{\textbf{PowerLaw}}&\multicolumn{3}{c}{\textbf{Cutoff PowerLaw}}      & \\
\hline

\textbf{Year}&\textbf{Index}&\textbf{TS Value}&\textbf{Index}&\textbf{CutOff (GeV)}&\textbf{TS Value}&\textbf{\boldmath{$\Delta$}TS} \\
\hline

2010   & 3.03 $\pm$ 0.03  &  463      & 2.48 $\pm$ 0.11 & 0.67 $\pm$ 0.17 & 467      & 4   \\
2014   & 3.19 $\pm$ 0.04  &  315      & 2.51 $\pm$ 0.07 & 0.43 $\pm$ 0.06 & 398      & 83  \\
2017   & 2.94 $\pm$ 0.07  &  512      & 2.13 $\pm$ 0.03 & 0.56 $\pm$ 0.05 & 538      & 26  \\
2021   & 2.99 $\pm$ 0.07  &  636      & 2.34 $\pm$ 0.03 & 0.59 $\pm$ 0.09 & 689      & 53  \\
\hline
\label{tab_model_pars}
\end{tabular}
\end{table}

The physical origin of the GeV flares remains unknown.
In PSR B1259-63, the neutron star
will cross the equatorial disk hosted by its companion Be star twice,
as an inclination angle exists between the disk and the orbital plane.
A bow-shaped shock region is generated
by the interaction between the stellar wind of the companion star
and the pulsar wind from the neutron star.
Theoretical models have been proposed to explain the GeV emission of PSR B1259-63 system.
In \citet{2020MNRAS.497..648C},
the GeV data are explained as a combination of the bremsstrahlung and
inverse Compton emission from the un-shocked and weakly shocked electrons of the pulsar wind,
assuming the pulsar wind is confined within a cone during the flares.
A high-speed ejecta that appeared as an extended X-ray structure, a clump,
moving away from PSR B1259-63 has been revealed in X-ray observation
\citep{2011ApJ...730....2P, 2015ApJ...806..192P, 2014ApJ...784..124K}.
The accelerated clump is ejected at every binary cycle near periastron passage
\citep{2019ApJ...882...74H}.
As suggested in model proposed in \citet{2021Univ....7..242C},
the bremsstrahlung emission from the clumps of the Be star wind entering the emission cone
can be used to explain the observed variability of GeV emission in the smaller time scale.

The reason for the increasing delay of the main GeV flux peaks is unknown and
this phenomena is discussed in previous literature.
The size of a clump or the lifetime of smaller clumps in the system
are thought to be a possible origin of this in \citet{2021Univ....7..242C}.
However, this effect may lead to a randomization of the delay time.
As we observed in our results, the peak time of the GeV flux (days after periastron)
shows a systematic non-linear trend of delay, which indicates a non-random origin.
It has been suggested that
A 0535 + 262, one of the best-studied Be/X-ray binaries,
has a precessing warped Be disk \citep{2013PASJ...65...83M}.
If the equatorial disk of Be star in PSR B1259-63 system has a slow precessing motion,
similarly to A 0535 + 262,
it may lead to a systematic delay of GeV flux in each periastron.
In this scenario,
the pulsar will enter the Be disk later each orbit,
if it is orbiting in the same direction as the motion of the precessing Be disk.
As a simple estimation,
we fitted the GeV peak time with a sinusoidal function (Figure \ref{fig_delay}),
leading to a period of $\sim$630,000 days, which might be linked with the precession.
If the trend of delay continues, in next periastron passage (30 June 2024)
the GeV peak will happen $\sim$150 days after periastron, around 27 November 2024.
However, multi-wavelength emission of PSR B1259-63 shows little delay between periastrons \citep{2021Univ....7..242C}.
If the precessing disk is at work,
we expect to see the delay of emission in other wavelength as well.
This is against the scenario and and further detailed studies are needed in the future.

\section{Conclusions} \label{sec_conclusions}

In this paper, we report
on \textit{Fermi}-LAT observations of PSR B1259-63 during its last four periastron passages.
We analyzed all the \textit{Fermi}-LAT public data
during its periastron passages in 2010, 2014, 2017 and 2021.
The long term observation of PSR B1259-63
allows its GeV $\gamma$-ray timing properties to be robustly characterized:
\begin{itemize}
  \item The presence of multiple GeV flares during each periastron passage is confirmed.
  \item A time delay of the main GeV flux peaks exists in its last four periastron passages.
\end{itemize}

We explored the possibility that above results are due to a slow precessing warped Be equatorial disk.
However, this scenario is not fully consistent with current multi-wavelength observations
and further studies are needed in the future.

\begin{acknowledgments}
This work is supported by the National Key R\&D Program of China (2016YFA0400800)
and the National Natural Science Foundation of China
under grants U2038101, U1838202, U1838201, 11733009, U1838104, U1938101, U1938103.
The authors acknowledge the anonymous referees for their important 
comments and suggestions.
\end{acknowledgments}

\vspace{5mm}
\facilities{\textit{Fermi}-LAT}

\software{}

\bibliography{ms}{}
\bibliographystyle{aasjournal}

\end{document}